\documentstyle[12pt]{article}
\begin{document}

\begin{center}
{\bf HEAVY BARYON MASSES\\}

\vspace*{1cm}
Agnieszka  Zalewska\\
{\it Institute of Nuclear Physics\\
ul. Kawiory 26a, Cracow 30055, Poland\\}

\vspace*{0.5cm}
and
\vspace*{0.5cm}

Kacper Zalewski\\
{\it Institute of Physics, Jagellonian University\\
ul. Reymonta 4, Cracow 30059, Poland\\
Institute of Nuclear Physics\\
ul. Kawiory 26a, Cracow 30055, Poland\\}
\end{center}

\vspace*{1.5cm}
\begin{abstract}
{\small Simple and plausible rules are used to correlate the masses of the
ground-state baryons containing single heavy ($b$ or $c$) quarks. A comparison
with the experimental data shows that the observed mass difference between the
$\Sigma_b$ and the $\Sigma^*_b$ is unexpectedly large. Predictions for the
masses of the yet undiscovered heavy baryons are given.}
\end{abstract}

\vspace*{1.5cm}
\section{Introduction}
The subject of this report are heavy baryons containing one heavy quark each.
Since the lifetime of the $t$-quark is too short for hadronization, the heavy
quark $Q$ is the $c$-quark, or the $b$-quark. From $SU(3)$ symmetry applied to
the light diquark in the baryon, or from the simple quark model, one expects
for each $Q$:
\begin{itemize}
\item A spin 1/2 sextet consisting of the isotriplet $\Sigma_Q$, the isodoublet
$\Xi'_Q$ and the isosinglet $\Omega_Q$;
\item A spin 3/2 sextet consisting of the isotriplet $\Sigma^*_Q$, the
isodoublet $\Xi^*_Q$ and the isosinglet $\Omega^*_Q$;
\item A spin 1/2 antitriplet consisting of the isosinglet $\Lambda_Q$ and the
isodoublet $\Xi_Q$.
\end{itemize}
Each of the states enumerated here should have excited states, but we limit our
discussion to the $2\times 8$ ground state isomultiplets. At present the
isomultiplets $\Sigma_c,\;\Omega_c\;\Lambda_c,\;\Xi_c,\;\Xi^*_c$ and
$\Lambda_b$ are reasonably well known$^{1,2,3}$. The isomultiplets
$\Xi'_c,\;\Omega^*_c,\;\Xi'_b,\;\Omega_b,\;\Xi_b,\;\Xi^*_b$ and $\Omega^*_b$
have not yet been observed. The $\Sigma^*_c$ has been observed$^4$, but with
poor statistics ($8$ events). For the $\Sigma_b$ and the $\Sigma^*_b$ only
preliminary data is available$^3$.

Rather surprisingly, almost any model yields "good" predictions for the masses
of these baryons. Thus e.g. Martin and Richard$^5$ quote ten predictions for
the mass of the $\Omega_c$, most of them made before the particle was
discovered, and the error nowhere exceeds $100$MeV, i.e. about $4$ per cent.
This makes one suspect that behind the detailed models there are some simple
rules, which are enough to get the good predictions and are so obvious that
they are explicitly, or implicitly, included in an approximate way in all
the models. Let us stress two advantages of identifying these simple rules.
They make it easy to distinguish at a glance the unexpected from the expected
in the experimental data. They also help to identify the "really good" models,
which can explain the small discrepancies between the predictions of the simple
rules and the data. A tentative set of such rules is formulated in the
following  section.

\section{Simpleminded rules for the baryon masses}

\begin{itemize}

\item {\bf Rule I} For $B =
\Sigma,\;\Xi',\;\Omega,\;\Xi^*,\;\Omega^*,\;\Lambda,\;\Xi$ the difference
$\delta_1 = M_{B_b} - M_{B_c}$ does not depend on $B$. Using the measured
values$^{1,3}$ of $\Lambda_b$ and $\Lambda_c$ one finds $\delta_1 = (3.353 \pm
0.016)$GeV. This (approximate!) rule is suggested by the heavy quark effective
theory. In order to estimate its uncertainty one can use the analogous rule for
mesons, which gives $M_{B_s} = M_B + M_{D_s} - M_D$. From the Particle Data
Group numbers$^1$ the right-hand-side is $5.380$ GeV, while according to a
recent measurement$^7$ $M_{B_s} = (5.370 \pm 0.03)$ GeV. Thus, to Rule I we
ascribe a systematic error of $\pm 10$ MeV on top of the experimental
uncertainty in the value of $\delta_1$.

\item {\bf Rule II} The mass differences between the adjacent isomultiplets in
the sextets are all equal. From the measured values of $M_{\Sigma_c}$ and
$M_{\Omega_c}$ one finds for this mass difference $\delta_2 = 0.127$ GeV. Since
the systematic error of this rule is bigger than the experimental uncertainty
of $\delta_2$, we have neglected this uncertainty. Rule II is suggested by the
quark model, where these mass differences are essentially due to the
replacement of a light quark ($u,d$) by the somewhat heavier $s$-quark. In
order to estimate the systematic error of Rule II, we looked on the deviations
of the experimental data from the predictions of the analogous rule for the
decuplet of light baryons and on the predictions of a model described by
Rosner$^6$, where this rule is only an approximation. Both suggest an error of
$\pm 10$ MeV.

\item {\bf Rule III} The hyperfine splittings, i.e. the mass differences
between the members of the spin 1/2 sextet and the corresponding members of the
spin 3/2 sextet are inversely proportional to the heavy quark masses. This is
again suggested by the heavy quark effective theory. From the measured masses
of $\Sigma_c,\;\Omega_c$ and$^8$ $\Xi^*_c$, using Rule II, we obtain for $Q =
c$ the splitting --- $\delta_{3c} = 63$ MeV and from that for $Q = b$ the
splitting --- $\delta_{3b} = 21$ MeV. For mesons the analogous rule works very
well, $(M_{B^*} - M_B)/(M_{D^*} - M_D) \approx \frac{1}{3}$ as expected. The
preliminary DELPHI data$^3$, however, give $M_{\Sigma^*_b} - M_{\Sigma_B} = (56
\pm 13)$ MeV. As discussed further this result is unexpected and when confirmed
can be of great importance for model builders. For the moment we ascribe to
Rule III the large systematic uncertainty $\pm 20$ MeV, which reduces the
discrepancy between the prediction of this rule and the preliminary data to
below two standard deviations.

\end{itemize}

Let us comment on the number of free parameters in our rules. We take from
experiment $\delta_1,\;\delta_2,\;\delta_{3c},\;\delta_{3b},\; M_{\Sigma_b},\;
M_{\Lambda_b}$ and $M_{\Xi_b}$ i.e. seven parameters. This is rather standard.
E.g. a typical quark model would have used $m_u = m_d,\;m_s,\;m_c,\;m_b$ and
three parameters in the potential (one should keep in mind the constant, which
is often not written explicitly).

\section{Tests}

There are three masses, which can be obtained using the rules given in the
preceding section and compared with already available experimental data.

For $\Sigma^*_c$ we find the mass $M_{\Sigma_c} + \delta_{3c} = (2.516 \pm
0.010)$ GeV in reasonably good agreement with the experimental result$^4$
$(2.530 \pm 0.007)$ GeV.

For $\Sigma_b$ we find $M_{\Sigma_b} = M_{\Lambda_b} + (0.168 \pm 0.005)$ GeV
in very good agreement with the DELPHI$^3$ result $(M_{\Lambda_b} + (0.173 \pm
0.009)$ GeV. Note that calculating the mass difference with respect to the
$\Lambda_b$ one reduces the error on both the theoretical prediction and the
experimental result$^3$.

For the $\Sigma^*_b$ we find a mass $M_{\Lambda_b} + (0.189 \pm 0.020)$ GeV,
while the DELPHI result is $M_{\Lambda_b} + (0.229 \pm 0.009)$ GeV. Due to the
large uncertainty ascribed to our prediction the discrepancy in standard
deviations is not outrageous, but we would like to stress that the potential
theoretical problem here is serious. Let us consider two ways of visualizing
this.

By hyperfine splitting (HFS) we understand the difference between the mass of
the heavy hadron, where the spin of the light component is parallel to the spin
of the heavy quark, and the mass of the corresponding heavy hadron, where these
spins are antiparallel. For the ratio of the HFS in the $b$-sector to the HFS
in the $c$-sector data give $\frac{1}{3}$ for mesons and $0.9$ for baryons This
large increase is not expected. The ratio of the HFS for baryons to the HFS for
mesons is below $0.5$ in the $c$-sector and about $1.2$ in the $b$-sector,
which is also unexplained.

\section{Predictions and conclusions}

Our predictions for the masses of the yet undiscovered baryons are:

\begin{eqnarray}
M_{\Xi'_c} & = & 2.580 \pm 0.010 \mbox{GeV}\\
M_{\Omega^*_c} & = & 2.770 \pm 0.010 \mbox{GeV}\\
M_{\Xi_b} & = & M_{\Lambda_b}  + (0.183 \pm 0.010)\mbox{GeV}\\
M_{\Xi'_b} & = & M_{\Lambda_b}  + (0.295 \pm 0.010)\mbox{GeV}\\
M_{\Omega_b} & = & M_{\Lambda_b}  + (0.422 \pm 0.010)\mbox{GeV}\\
M_{\Xi^*_b} & = & M_{\Lambda_b}  + (0.316 \pm 0.020)\mbox{GeV}\\
M_{\Omega^*_b} & = & M_{\Lambda_b}  + (0.443 \pm 0.020)\mbox{GeV}
\end{eqnarray}
The errors on these predictions are estimated using analogies only. They
correspond to one standard deviation rather than to the maximum conceivable
uncertainties. The predicted values may be interpreted as predictions of one
more model, however, they may be also useful as bench marks. If a model gives
predictions less good than these, their agreement with experiment within some
stated uncertainties may be used as an argument for the soundness of the model,
but the predictions as such are of little interest. If the predictions of a
 model are better than those from our simple rules, it is interesting to
identify  the physics behind these improvements. A possible problem is the
$\Sigma^*_b,\;\Sigma_b$ HFS as compared with the analogous splittings in the
$c$-sector and in the meson sectors. On should keep in mind, however, that the
experimental data suggesting that there is a problem is still preliminary.

\section{Acknowledgements}
This work has been supported in part by the KBN grants 2P302-112-06 and
2P302-076-07.

\section{References}

\noindent 1. Particle Data Group, {\it Phys. Rev.} {\bf D50} (1994) 1173.\\
2. F. Dropmann, Rencontre de Moriond (1995) in print. \\
3. D. Bloch, Report at the EPS Conference Brussels (1995) in print.\\
4. V. A. Amosov et al., {\it Pis'ma Zh. Eksper. Theor. Fiz.} {\bf 58} (1993)
241.\\
5. A. Martin and J.-M. Richard, {\it Phys. Letters} {\bf B355} (1995) 345.\\
6. J.L. Rosner, {\it Phys. Rev.} {\bf D52} (1995) 6461.\\
7. F. Abe et al., {\it Phys. Rev.} {\bf D53} (1996) 241.\\
8. P. Avery et al., {\it Phys. Rev. Letters} {\bf 75} (1995) 4364.

\end{document}